\documentclass{desyproc}

\DeclareMathOperator{\MeV}{MeV}

\DeclareMathOperator{\kpc}{kpc}
\DeclareMathOperator{\kton}{kton}
\newcommand{\oxy}{$^{16}$O\,}
\usepackage{gensymb}
\begin{document}
\title{Axion emission and detection from a Galactic supernova
}

\author{{\slshape Pierluca Carenza$^1$, Giampaolo Co' $^{2,}{^3}$, Tobias Fischer$^4$, Maurizio Giannotti$^5$, Alessandro Mirizzi$^{1,}{^6}$\footnote{Speaker}, Thomas Rauscher$^{7,}{^8}$}\\[1ex]
$^1$Dipartimento Interateneo di Fisica ``Michelangelo Merlin'', Universit\`a degli Studi di Bari, Bari, Italy\\
$^2$Dipartimento di Matematica e Fisica ``E. De Giorgi'', Universit\`a del Salento, Lecce, Italy\\
$^3$Istituto Nazionale di Fisica Nucleare - Sezione di Lecce, Lecce, Italy\\
$^4$Institute of Theoretical Physics, University of Wroc{\l}aw, 50-204 Wroc{\l}aw, Poland\\
$^5$Physical Sciences, Barry University, 11300 NE 2nd Ave., Miami Shores, FL 33161, USA\\
$^6$Istituto Nazionale di Fisica Nucleare - Sezione di Bari, Bari, Italy\\
$^7$Department of Physics, University of Basel, Basel, Switzerland\\
$^8$Centre for Astrophysics Research, University of Hertfordshire, Hatfield, UK\\}

\contribID{Lindner\_Axel}

\confID{20012}  
\desyproc{DESY-PROC-2018-03}
\acronym{Patras 2018} 
\doi  

\maketitle

\begin{abstract}
A Galactic supernova (SN) axion signal would be detected in a future neutrino Mton-class
water Cherenkov detector, such as the proposed Hyper-Kamiokande
in Japan. The main detection channel for axions is
 absorption on the oxygen nuclei in the water. The subsequent oxygen de-excitation leads to a potentially detectable gamma signal. In this contribution we present a calculation of the SN axion signal and discuss its detectability in Hyper-Kamiokande.
\end{abstract}

\section{Introduction}

Low-mass axions can be copiously produced in a core-collapse supernova (SN) affecting the neutrino burst. In fact, bounds on axions have been placed studying the neutrino signal from SN 1987A. In particular, in \cite{engels} it was pointed out that  a SN axion burst could produce an observable signal in a neutrino water Cherenkov detector by oxygen absorption. The following oxygen de-excitation would produce a photon signal. Currently it has been proposed a future  Mton-class neutrino water Cherenkov detector, Hyper-Kamiokande, in Japan. Motivated by this exciting situation, we find it worthwhile to take a fresh look at the possibility of detecting a SN axion burst.  In this contribution we present preliminary results of an updated calculation of the SN axion signal in  Hyper-Kamiokande. We refer the interested reader to \cite{master,prep} for further details.

\section{SN axion flux}

Axions are produced in a SN enviroment via nucleon-bremsstrahlung $NN\rightarrow NNa$. The axion-nucleon coupling constant $g_{aN}$ depends on the Peccei-Quinn scale $f_{a}$ as $g_{aN}=C_{N}m_{N}/f_{a}$, where $N=p,n$, $m_{N}$ is the nucleon mass and $C_{N}$ is a model-dependent factor.
We computed the axion and neutrino fluxes through a SN simulation developed by the ``Wroclaw Supernova Project''
based on a spherically symmetric core-collapse SN model, using the AGILE-BOLTZTRAN code~\cite{fischer1,fischer2}. 
We consider two representative cases:
\begin{itemize}
\item Weakly-interacting axions with $g_{ap}=9\times10^{-10}$ and $g_{an}=0$. In this case, axions are in a free-streaming regime and they drain energy from the SN core, suppressing the neutrino fluxes.
\item Strongly interacting axions with $g_{ap}=g_{an}=10^{-6}$. Axions are in a trapping regime, so they do not reduce  the neutrino fluxes. However, they are emitted from a last-scattering surface, the axionsphere. In this case their flux can be larger than the neutrino ones.
\end{itemize}

\section{Axion-\oxy absorption cross section}
\label{sec:cross}

Axions can be detected in a water Cherenkov detector via axion-\oxy absorption, $a ^{16}{\rm O}\rightarrow{}^{16}{\rm O}^{*}$, revealing the oxygen  decays in photons. The absorption cross section is evaluated starting from the following axion-nucleon interaction Lagrangian \cite{engels}
\[
\begin{split}
&\mathcal{L}=\frac{1}{2f_{a}} \bar{\Psi}_{N}\gamma^{\mu}\gamma^{5}(C_{0}+C_{1}{\bf{\tau}}_{3})\Psi_{N}\partial_{\mu}a\;;\\
&C_{0}=\frac{1}{2}(C_{p}+C_{n})\;;\\
&C_{1}=\frac{1}{2}(C_{p}-C_{n})\;;
\end{split}
\]
where $\Psi_{N}$ is the nucleon spinor and ${\bf{\tau}}_{3}$ is a Pauli matrix.
The obtained cross section is 
\begin{equation}
\sigma=\frac{4\pi^{2} E_{p}}{f_{a}^{2}}|\langle J^{P}||L_{j,0}||0^{+}\rangle|^{2}\;;
\label{eq:cross}
\end{equation}
where 
\[
\begin{split}
&L_{j,0}=\frac{i}{p}\int d^{3}r\, \partial_{i} (j_{j}(pr)Y_{j,0}(\Omega))J^{i}(\bf{r})\;;\\
&J^{i}({\bf{r}})=\bar{\Psi}_{N}({\bf{r}})\gamma^{i}\gamma^{5}(C_{0}+C_{1}{\bf{\tau}}_{3})\Psi_{N}({\bf{r}})\;.
\end{split}
\]

The reduced matrix element in Eq.~(\ref{eq:cross}) is calculated between the \oxy ground state, $|0^{+}\rangle$, and the $^{16}{\rm O}^{*}$ excited state, $|J^{P}\rangle$. The angular momentum and parity of the excited oxygen are fixed by conservation laws. Therefore $J$ is the axion angular momentum and $P=(-1)^{J+1}$. The cross section in Eq.~(\ref{eq:cross}) is explicitly computed with the Random Phase Approximation. Particle- and $\gamma$-emissions from the excited \oxy were computed using transmission coefficients from the SMARAGD Hauser-Feshbach code~\cite{x1}. Two-particle emission was included. The obtained total $\gamma$-ray and particle spectra were folded with the detector properties to obtain the expected event number.

\section{Axion events}
As reference detector we consider Hyper-Kamiokande, a next-generation water Cherenkov detector, with $M=374\kton$ of fiducial mass.
The detected neutrino (or axion) events in the proposed detector are calculated as
\[
N_{\rm ev}=F\otimes\sigma\otimes \mathcal{R}\otimes\mathcal{E}\;;
\]
where $F$, the neutrino (or axion) flux, is convoluted with the cross section $\sigma$ in the detector, the detector energy resolution, $\mathcal{R}$ and the detector efficiency $\mathcal{E}$. 
We assume $\mathcal{E}=1$ above the energy threshold ($E_{\rm th}=5\MeV$) and the energy resolution is the same of the Super-Kamiokande detector. 

With the given energy threshold, the majority of the detectable axion signal falls in the range $5-10\MeV$. Therefore we restrict our attention to this energy window. 
The neutrino interaction reactions in the detector are a background for the axion detection. In particular, one has to consider the following channels:
inverse beta decay  (IBD), $\bar{\nu}_{e}p\rightarrow ne^{+}$; elastic scattering (ES), $\nu e^{-}\rightarrow
\nu e^{-}$; charged and neutral current $\nu$-\oxy nuclei interactions (O-CC and O-NC). The number of free-streaming axion and neutrino events in the range $5-10\MeV$ is shown in Tab.~1. The huge neutrino background dominates the axion signal. However,
the axion detectability can be enhanced doping the detector with gadolinium (Gd) to tag the IBD events. This possibility is currently being realized in Super-Kamiokande. We assume that it would occur also for Hyper-Kamiokande.
Gd has a large neutron capture cross section and, after a neutron capture, emits a cascade of photons with a total energy of $8\MeV$. 
The coincidence detection of the positron and photon signals tags the IBD events.
We will assume $90\%$ tagging efficiency as quoted in \cite{vagins}. 
The ES signal can be reduced through a directional cut. Indeed the scattered electrons preserve the incident neutrino direction.
Then the majority of ES events (about $95\%$) is contained in a $40^{\degree}$ cone, making possible a reduction of this background by means of a directional cut which eliminates also the $12\%$ of the events in the other channels  \cite{tomas}.
The number of events with the background reduction is shown in Tab.~1. We observe that the background reduction makes it  possible to detect axions at less than $2\sigma$ for a SN at $d=1\kpc$. 

Furthermore, a new calculation of the axion bremsstrahlung production in a SN \cite{Chang:2018rso}, shows that the SN axion flux should be about 20 times lower than the one we used until now. Since the flux in the free-streaming is proportional to $g_{ap}^{2}$, to obtain the same luminosity we should use a coupling constant larger by a factor of $\sqrt{20}$. Then the number of axion events is 20 times larger than our previous estimate [Tab.~1]. In this case, the axion signal would emerge at $\sim 28\sigma$ for a SN at $1\kpc$ and at $\sim 3\sigma$ for a SN at $10\kpc$.

We also calculated the events in the trapping regime, obtaining the results in Tab.~2. In this regime the axion signal dominates the neutrino background.

\begin{table}[t] 
\centerline{\begin{tabular}{|c|c|c|c|}
\hline
 $g_{ap}=9\times10^{-10}, g_{an}=0$&  &&  \\
\hline
Interaction&Events & BKG RED&NEW FLUX  \\
\hline
a-O&270& 238&$4.76\times10^{3}$\\
IBD &$1.99\times10^{5}$ &$1.75\times10^{4}$&$1.75\times10^{4}$\\
ES & $3.53\times10^{4}$&$1.77\times10^{3}$&$1.77\times10^{3}$\\
O-CC &$1.76\times10^{3}$&$1.55\times10^{3}$&$1.55\times10^{3}$\\
O-NC &$9.21\times10^{3} $&$8.10\times10^{3}$&$8.10\times10^{3}$\\
 \hline
\end{tabular}}
\label{tab:completo}
\caption{Number of events without background reduction (first column), number of events with
 background reduction (second column) and number of events with the flux correction factor (third column) in the range $[5;10]\MeV$ for a SN at $d=1\kpc$ and a detector mass $M=374\kton$.}
\end{table}

\begin{table}[h]
\centerline{\begin{tabular}{|c|c|}
\hline
 $g_{ap}=g_{an}=10^{-6}$ & \\
\hline
Interaction&Events \\
\hline
a-O& $2.73\times10^{5}$\\
IBD &$2.85\times10^{3}$\\
ES & 522\\
O-CC &24\\
O-NC &116\\
 \hline
\end{tabular}}
\label{tab:trap}
\caption{Number of events with
 background reduction in the range $[5;10]\MeV$ for a SN at $d=10\kpc$, a detector mass $M=374\kton$ and a coupling constant  $g_{ap}=g_{an}=10^{-6}$.}
\end{table}

\section{Conclusions}
We evaluated the Hyper-Kamiokande potential to detect the axion burst associated with a Galactic SN event. We found that axions  
in the free-streaming regime would be potentially detectable if a careful reduction of the neutrino background is  performed.  On the other hand, in the trapping regime the axion signal would dominate over the neutrino one, being easily detectable. Therefore a Galactic SN explosion would be a once in a lifetime opportunity for detecting axions.

\section*{Acknowledgments}
The work of A.M. is supported by the Italian Istituto Nazionale di Fisica Nucleare (INFN) through the ``Theoretical Astroparticle Physics'' project and by Ministero dell' Istruzione, Universit\`a e Ricerca (MIUR). Futhermore, A.M. acknowledges support from the Alexander von Humboldt Foundation for his participation in the Patras Workshop 2018.
T.F. acknowledges support from the Polish National Science Center (NCN) under grant numbers UMO-2016/23/B/ST2/00720. The supernova simulations were performed at the Wroclaw Center for Scientific Computing and Networking (WCSS). 
T.R. is partially supported by the  EU COST Action CA16117 (ChETEC).


\begin{footnotesize}
\begin{footnotesize}

\end{footnotesize}

\end{footnotesize}



\begin{thebibliography}{99}
%

\bibitem[1]{engels}
  J.~Engel {\it et al.},
  Phys.\ Rev.\ Lett.\  {\bf 65} (1990) 960.
  
\bibitem[2]{master}
P.~Carenza, ``Axion emission and detection from a galactic supernova,''
Master Thesis (2018).

\bibitem[3]{prep}
  P.~Carenza, G.~C\`o, T.~Fischer, M.~Giannotti, A.~Mirizzi, and T.~Rauscher,
  in preparation.

\bibitem[4]{fischer1}
M.~Liebendoerfer {\it et al.},
  Astrophys.\ J.\ Suppl.\  {\bf 150} (2004) 263
  [astro-ph/0207036].
  
  \bibitem[5]{fischer2}
  T.~Fischer {\it et al.},
  Phys.\ Rev.\ D {\bf 94} (2016) no.8,  085012
  [arXiv:1605.08780 [astro-ph.HE]].
  
  
  \bibitem[6]{x1} T.~Rauscher, code SMARAGD, version v0.11.0sn (2018).
    
  \bibitem[7]{vagins}
  J.~F.~Beacom and M.~R.~Vagins,
  Phys.\ Rev.\ Lett.\  {\bf 93} (2004) 171101
  [hep-ph/0309300].
  
  \bibitem[8]{tomas}
  R.~Tomas {\it et al.},
  Phys.\ Rev.\ D {\bf 68} (2003) 093013
  [hep-ph/0307050].
  
  
\bibitem[9]{Chang:2018rso}
  J.~H.~Chang, R.~Essig and S.~D.~McDermott,
  arXiv:1803.00993 [hep-ph].

  
\end{thebibliography}
\end{document}